\documentclass[twocolumn,showpacs,preprintnumbers,amsmath,amssymb,prx,nofootinbib,aps]{revtex4-2}

\usepackage{graphicx}
\usepackage{dcolumn}
\usepackage{bm}
\usepackage{color}
\usepackage{rotating}
\usepackage{ulem} 
\graphicspath{{figures/}}

\begin{document}
\newcommand\blfootnote[1]{%
	\begingroup
	\renewcommand\thefootnote{}\footnote{#1}%
	\addtocounter{footnote}{-1}%
	\endgroup}
    
\newcommand{\sh}[1]{\textcolor{blue}{#1}}

 \newcommand{\FigCap}[1]{\textbf{#1}}	
\newcommand{\red}{\textcolor{red}}
\newcommand{\blue}{\textcolor{blue}}
\newcommand{\green}{\textcolor{green}}

\title{Pseudo Point Nodal Superconducting Gap in Spin-Triplet UTe$_2$}

\setlength{\textfloatsep}{10pt plus 1pt minus 2pt}

\author{S.\,Hosoi$^{1,*}$}
\author{K.\,Imamura$^{1,2}$}
\author{M.M.\,Bordelon$^1$}
\author{E.D.\,Bauer$^1$}
\author{S.M.\,Thomas$^1$}
\author{F.\,Ronning$^1$}
\author{P.F.S.\,Rosa$^1$}
\author{R. Movshovich$^1$}
\author{I.\,Vekhter$^3$}
\author{Y.\,Matsuda$^{1,\dagger}$}

{\let\thefootnote\relax\footnote{corresponding authors: \\
$^{*}$shosoi@lanl.gov, ~~$^{\dagger}$matsuda@lanl.gov}}

\affiliation{$^1$Los Alamos National Laboratory, Los Alamos, NM 87545, USA}
\affiliation{$^2$Department of Advanced Materials Science, University of Tokyo, Chiba 277-8561, Japan} 
\affiliation{$^3$
Department of Physics and Astronomy, Louisiana State University, Baton Rouge, LA 70803-4001, USA}

\maketitle
{\bf 
The unconventional superconductor $\mathrm{UTe}_2$ represents a rare example of spin-triplet pairing with potentially topologically protected quantum states. However, conflicting reports on its gap structure---particularly regarding point nodes---have hindered understanding of the order parameter symmetry and topological properties. Here, we report high-resolution thermal conductivity measurements on high-quality $\mathrm{UTe}_2$ single crystals down to $\sim$50\,mK that definitively resolve the gap anisotropy through bulk directional transport. The  $b$-axis thermal conductivity $\kappa_b/T$ exhibits negligible residual conductivity at $T \to 0$ and its temperature dependence is consistent with a small superconducting energy gap along the $b$-axis. Under magnetic fields, the residual $\kappa_b/T$ exhibits weak field-induced enhancement. Remarkably, a threshold field emerges at low fields for $\boldsymbol{H} \parallel a$, characterized by a kink that signals a change in the quasiparticle transport normal to the field. Below the threshold, $\kappa_b/T$ remains isotropic for all field orientations, whereas strong anisotropy between transport along and normal to the field develops above it. These signatures strongly suggest that $\mathrm{UTe}_2$ exhibits a fully gapped state with a pseudo point-nodal structure---where gap minima approach but never reach zero. We estimate the minimal gap $\Delta_{\rm min}/\Delta_0 \sim 0.1$ along the $b$-axis, where $\Delta_0$ is the characteristic superconducting gap. This highly unusual gap structure provides crucial insights into the exotic pairing mechanisms and topology of this spin-triplet superconductor, excluding non-unitary mixing of pairing symmetries.}

Spin-triplet superconducting states have garnered significant attention due to their unconventional pairing mechanism and potential to host topologically protected Majorana zero modes, which offer a promising platform for fault-tolerant topological quantum computation \cite{nayak2008non}. While liquid $^3$He exemplifies spin-triplet superfluidity, identifying solid-state analogs remains a central challenge in condensed matter physics. Recently, the heavy fermion superconductor UTe$_2$ has emerged as an exceptionally promising candidate for realizing spin-triplet superconductivity \cite{ran2019nearly}. This uranium-based compound exhibits a constellation of anomalous superconducting properties that collectively provide compelling evidence for unconventional pairing mechanisms. Most notably, UTe$_2$ displays extraordinarily large upper critical fields $H_{c2}$ that substantially exceed the Pauli paramagnetic limit \cite{ran2019nearly,aoki2022unconventional}. The superconducting phase diagram under both magnetic field and pressure reveals additional complexity through the existence of multiple distinct superconducting phases \cite{ran2019extreme,knafo2021comparison,wu2025quantum,braithwaite2019multiple,thomas2020evidence}, indicating highly unusual pairing states.

Critically, the unconventional superconducting state in UTe$_2$ emerges from a paramagnetic normal state at ambient pressure, distinguishing it from uranium-based compounds such as UGe$_2$ \cite{saxena2000superconductivity}, URhGe \cite{aoki2001coexistence}, and UCoGe \cite{PhysRevLett.99.067006}, where superconductivity coexists with ferromagnetic order. Moreover, The Fermi surface structure of UTe$_2$ appears to be simpler than that of many other heavy fermion superconductors, with fewer sheets and less complex topology \cite{aoki2022unconventional,eaton2024quasi,PhysRevLett.123.217001,xu2019quasi}. These characteristics establish UTe$_2$ as an ideal model system for investigating spin-triplet superconductivity.

The comprehensive characterization of the superconducting gap structure in UTe$_2$ represents a critical prerequisite for understanding the Cooper pair formation mechanism and determining its topological nature.  However, despite extensive experimental investigations, the precise gap structure of UTe$_2$ remains unresolved. The challenge is further amplified by pronounced surface-bulk electronic structure disparities, most notably manifested through the emergence of a surface charge density wave (CDW) instability that does not occur in the bulk~\cite{aishwarya2023magnetic,talavera2025surface,kengle2024absence}. The surface-bulk dichotomy presents systematic challenges for interpreting surface-sensitive spectroscopic techniques, including scanning tunneling microscopy (STM) \cite{gu2025pair} and tunnel junction measurements \cite{li2025observation,yoon2024probing}.  Additionally, magnetic penetration depth measurements \cite{ishihara2021chiral,bae2021anomalous,iguchi2022microscopic}  may be influenced by the sample edge \cite{iguchi2022microscopic} or anomalous surface normal state \cite{bae2021anomalous}. We also note that, since low-energy quasiparticle excitations in UTe$_2$ are particularly sensitive to disorder, ultraclean crystals are required, whereas lower quality samples obscure intrinsic gap features.

 \begin{figure}[t]
\includegraphics[width=\columnwidth]{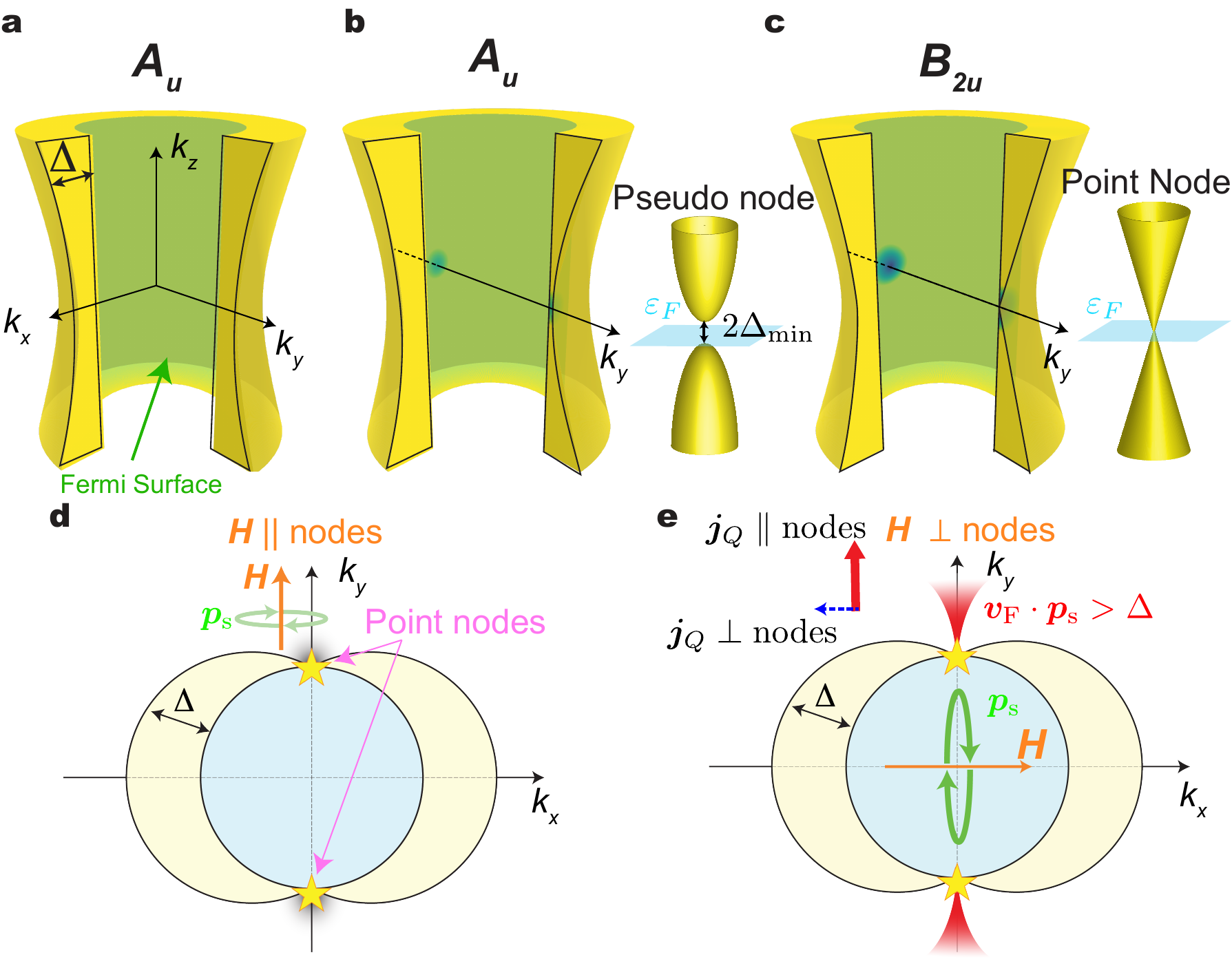}
\caption{\textbf{Superconducting gap structure and Doppler shift:} \textbf{a}, \textbf{b}, \textbf{c}: Superconducting gap structure $\Delta$ (yellow) on a 2D cylindrical Fermi surface (yellow-green) for $A_u$ (\textbf{a} and \textbf{b}) and $B_{2u}$ (\textbf{c}) symmetries. \textbf{a}: Full gap opens uniformly. \textbf{b}: At pseudo point nodes, the gap amplitude approaches zero but remains finite. The accompanying figure (right) shows the quasiparticle dispersion in the vicinity of the gap minimum. \textbf{c}:  The $B_{2u}$ state hosts point nodes (dark green) along the $b$-axis ($k_y$ direction). The Dirac cone emerges where the gap amplitude vanishes at a point node. \textbf{d,e}: Schematic representation of Doppler shift effects under magnetic field $\boldsymbol{H}$ applied perpendicular and parallel to point nodes, respectively. The supercurrent velocity $\boldsymbol{v}_s$ circulating around vortex cores is oriented perpendicular to $\boldsymbol{H}$. \textbf{d}: For $\boldsymbol{H} \parallel$ nodes, Doppler-shifted quasiparticle excitations are suppressed due to $\boldsymbol{v}_{\rm F}\cdot\boldsymbol{p}_{\rm s} = 0$. \textbf{e}: For $\boldsymbol{H} \perp$ nodes , the Doppler shift energy $\boldsymbol{v}_{\rm s}\cdot\boldsymbol{p}_{\rm s}$ exceeds the superconducting gap magnitude $\Delta$, thereby activating low-energy quasiparticle excitations in the nodal directions (red-shaded region). When thermal current $\boldsymbol{j}_Q$ is applied parallel to the nodal directions ($\boldsymbol{j}_Q \parallel$ nodes), thermal conductivity exhibits selective sensitivity to nodal excitations, while it becomes insensitive when $\boldsymbol{j}_Q \perp$ nodes.
} 
\label{fig:Doppler}
\end{figure}

Thermodynamic measurements under uniaxial strain \cite{girod2022thermodynamic} and ultrasound measurement \cite{theuss2024single}, combined with polar Kerr effect spectroscopy \cite{ajeesh2023fate} and muon spin relaxation measurements \cite{azari2023absence}, have provided  evidence against bulk chiral superconductivity. Accordingly, four single-component irreducible representations within the $D_{2h}$ point group characterize potential spin-triplet pairing states: $A_u$ (fully gapped, Figs.\,1a and 1b), $B_{1u}$ (point nodes along the crystallographic $c$-axis), $B_{2u}$ (point nodes along the $b$-axis, Fig.\,1c), and $B_{3u}$ (point nodes along the $a$-axis).

A definitive determination of the superconducting order parameter requires directionally-resolved, bulk-sensitive experimental probes capable of directly accessing nodal quasiparticle excitations at ultra-low temperatures \cite{matsuda2006nodal}. In contrast to specific heat, thermal conductivity preferentially samples itinerant quasiparticles with the Fermi velocity $\boldsymbol{v}_{\rm F}$ parallel to the thermal current $\boldsymbol{j}_Q$. Since in anisotropic superconductors at low energies such quasiparticles are confined to the regions when the gap is minimal, the directional dependence of thermal transport reflects the underlying nodal structure. Notably, thermal conductivity remains immune to extrinsic Schottky contributions, permitting measurements at sufficiently low temperatures to unambiguously differentiate nodal and gapped states. Most significantly, as the magnetic field also serves as a directional probe that excites quasipartcles through the Doppler shift of the quasiparticle spectra, the combination of magnetic field orientation with thermal conductivity measurements constitutes a powerful probe for superconducting gap determination, enabling precise mapping of nodal directions (Figs.\,1d and 1e).

Thermal conductivity studies in UTe$_2$ have established the absence of point nodes along the $a$ direction through the $a$-axis thermal conductivity ($\kappa_a$) measurements ({\boldmath $j$}$_Q \parallel a$)\cite{suetsugu2024fully,hayes2025robust}. In contrast, the conclusions regarding point nodes along the $b$ axis remain controversial. This discrepancy arises because $\kappa_a$ measurements probe mostly the quasiparticles with the Fermi velocity along the $a$-axis, and are much less sensitive to the the excitations in the vicinity of the potential $b$-axis nodes (Fig.\,1e). While $b$-axis thermal conductivity ($\kappa_b$) measurements ({\boldmath $j$}$_Q \parallel b$) would directly probe the $b$-axis nodal structure, such experiments were previously precluded by insufficient sample sizes along the $b$ direction, preventing definitive determination of the complete nodal topology.

\begin{figure*}[t]
    \label{fig:TCzero}	
    \includegraphics[width=0.85\linewidth]{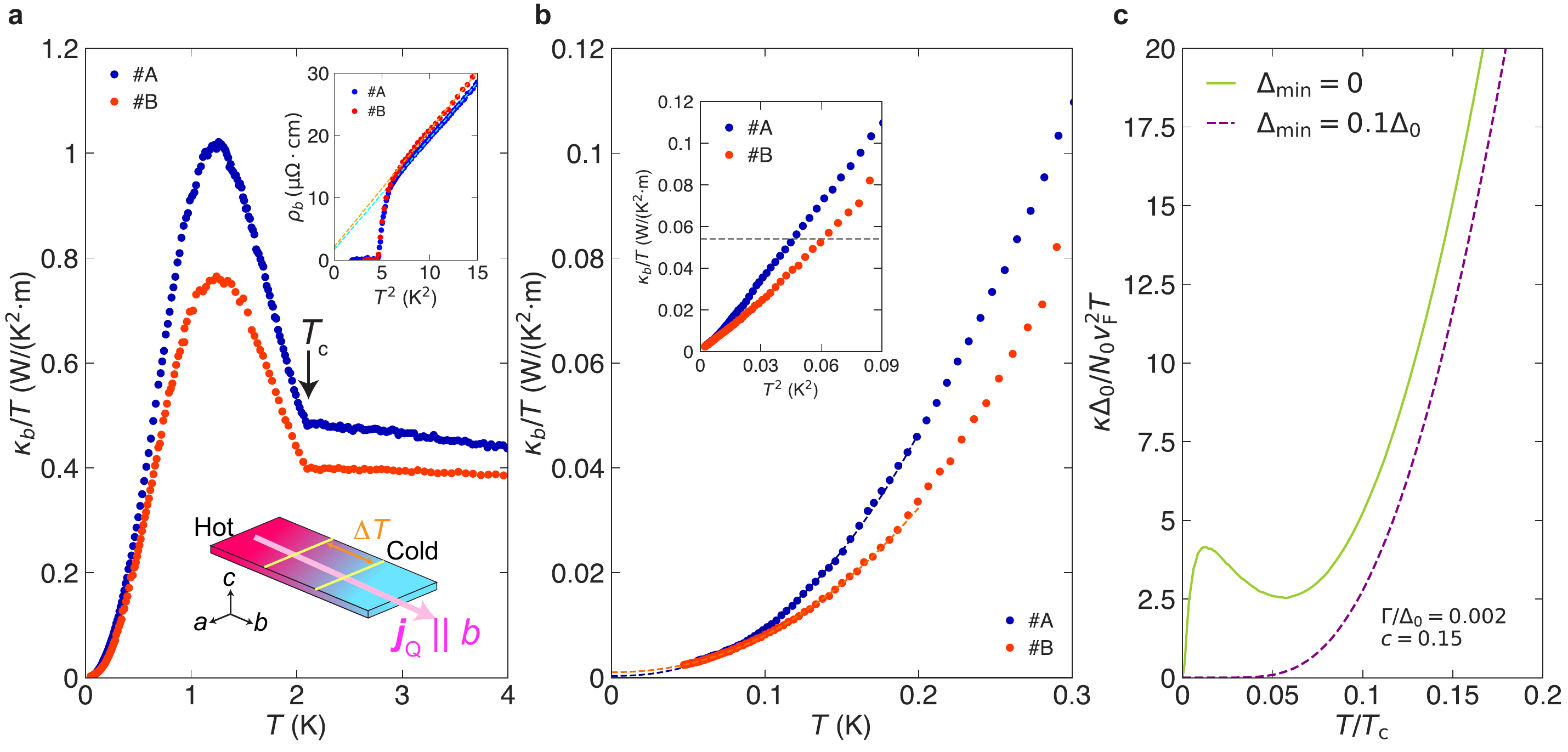}
\caption{\textbf{Thermal conductivity in zero magnetic field:} \textbf{a}, The main panel shows the $b$-axis thermal conductivity divided by temperature, $\kappa_b/T$, in zero field for sample \#A ($RRR \approx$410) and \#B ($RRR \approx$350). Both samples exhibit a sharp discontinuity at $T_c$ (arrow). The lower inset illustrates the experimental configuration for thermal conductivity measurements with thermal current {\boldmath $j$}$_Q$ applied along the crystallographic $b$-axis. The upper inset shows electrical resistivity along the $b$-axis in zero field plotted as a function of $T^2$ for \#A and \#B. $\rho_b(T)$ is well fit by $\rho_b(T)=\rho_b^0+AT^2$ (orange and cyan dashed line, respectively). \textbf{b}, $\kappa_b/T$ in zero field for \#A and \#B at very low temperatures. The inset shows $\kappa_b/T$ vs. $T^2$. The gray dashed line indicates the residual thermal conductivity expected for a superconductor with line nodes in the gap structure. \textbf{c}, Calculated temperature dependence of $\kappa/T$ for point nodes ($\Delta_{\rm min}=0$) and pseudo point nodes ($\Delta_{\rm min}=0.1\Delta_0$) , with heat current parallel to the nodal direction. 
}
\end{figure*}

Here, we measured the thermal conductivity $\kappa_b$ along the $b$-axis using molten-salt-grown ultraclean single crystals~\cite{PhysRevMaterials.6.073401} (lower inset of Fig.~2a). The upper inset of Fig.~2a shows the resistivity along the $b$-axis, $\rho_b$, plotted as a function of $T^2$ for samples \#A and \#B. The resistivity follows standard Fermi-liquid behavior, described by $\rho_b(T) = \rho_b^0 + AT^2$. This behavior is consistently observed under magnetic field $\boldsymbol{H} \parallel a$ down to very low temperatures (Supplementary Fig.\,1). We estimated residual resistivities of $\rho_b^0 \approx 1.72$~$\mu\Omega$cm for sample \#A and $\approx 2.12$~$\mu\Omega$cm for sample \#B in zero field. These values correspond to residual resistivity ratios of $RRR \equiv \rho_b(300\,\text{K})/\rho_b^0 \approx 410$ and $\approx 350$ for samples \#A and \#B, respectively.

We first examine the zero-field thermal conductivity.  We confirmed the Wiedemann-Franz law, providing stringent validation of thermal conductivity measurement accuracy (Supplementary Fig.\,2).  The main panel of Fig.~2a depicts the temperature dependence of $\kappa_b$. At the superconducting transition temperature $T_c=2.1$\,K, $\kappa_b/T$ exhibits a pronounced discontinuity, followed by a pronounced maximum at around 1.2\,K. The enhancement below $T_c$ for \#B (with lower $RRR$) is slightly suppressed compared to \#A. This enhancement arises from the dramatic increase in quasiparticle mean free path, which results from suppressed electron--electron inelastic scattering following superconducting gap opening. 

 The temperature dependence of $\kappa_b/T$ at very low temperatures for both sample\#A and \#B is shown in Fig.\,2b. Similar to $\kappa_a/T$ \cite{suetsugu2024fully,hayes2025robust}, $\kappa_b/T$ decreases monotonically upon cooling.  As shown in the inset of Fig.~2b, $\kappa_b/T$ exhibits a nearly $T^2$ dependence.  The thermal conductivity in the superconducting state comprises quasiparticle and phonon contributions. As discussed below, our results strongly indicate a substantial quasiparticle component in the measured thermal conductivity at low temperatures. For superconductors with line nodes, the low temperature residual thermal conductivity has a universal, impurity-independent, slope, which for UTe$_2$ is estimated to be $\kappa_{00}/T \approx 0.054$~W/K$^2$m (gray dashed line in the inset of Fig.~2b)~\cite{suetsugu2024fully}. The extrapolation of $\kappa_b/T$ to zero temperature (red and blue dotted lines in the main panel of Fig.\,2b) yields a vanishing intercept within experimental accuracy, indicating negligibly small residual thermal conductivity, with an upper bound estimate of $\kappa^0_b/T = 0.0017$~W/K$^2$m for both \#A and \#B.  Thus  $\kappa^0_b/T$ is significantly smaller than $\kappa_{00}/T$, similar to $\kappa_a/T$ \cite{suetsugu2024fully,hayes2025robust}. These results exclude line-nodal gap structures compatible with $D_{\mathrm{2h}}$ symmetry under weak spin-orbit coupling, constraining the pairing to either a full gap or point nodes as expected in the strong spin-orbit coupling regime.
 

\begin{figure*}[t]
    \label{fig:ER}
	\includegraphics[clip,width=0.8\linewidth]{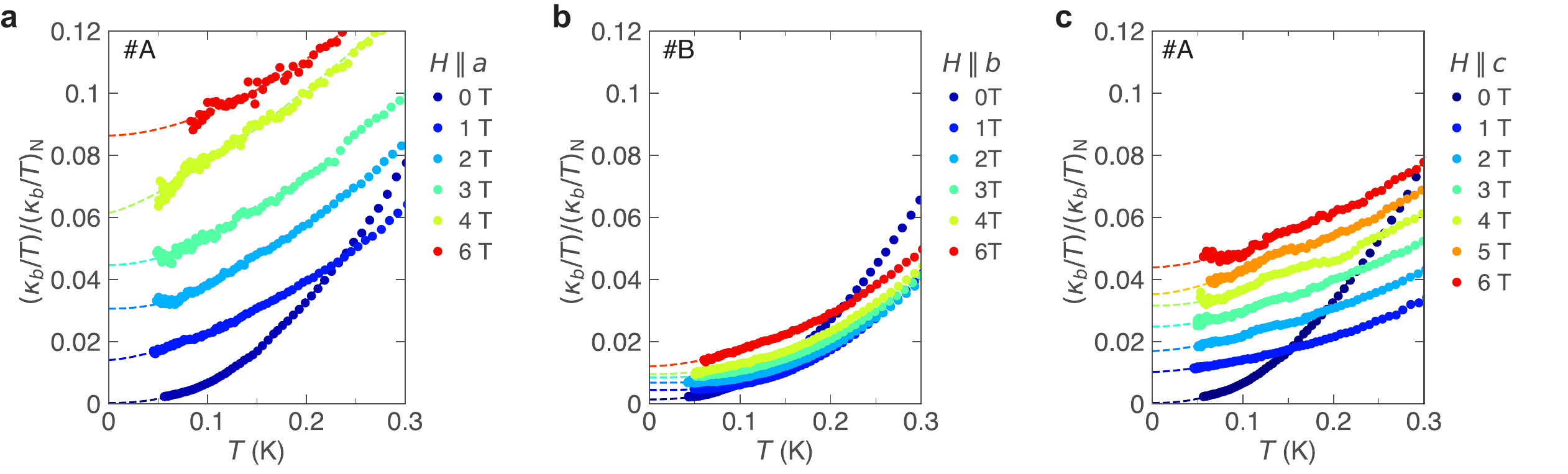}
	\caption{\textbf{Low temperature thermal conductivity under magnetic fields along each crystallographic axis:} \textbf{a,b,c}, Temperature dependence of low-temperature thermal conductivity along the  $b$-axis under magnetic fields applied parallel to the $a$-axis (\textbf{a}), $b$-axis (\textbf{b}), and $c$-axis (\textbf{c}), respectively. The dashed lines represent polynomial fitting results.}
\end{figure*}

Temperature dependence of $\kappa_b/T$ in zero field provides pivotal information regarding the presence or absence of point nodes along the $b$-axis~\cite{hayes2025robust}. For the heat current along the point nodes, the residual term in the thermal conductivity is finite only if the impurity scattering rate, $\Gamma$, exceeds a threshold value, and vanishes otherwise~\cite{hirschfeld_consequences_1988}. The threshold value is finite for any scattering phase shift, $\delta$, away from the unitary limit ($c=\cot\delta =0$). Therefore, the near absence of the residual linear term observed here is strongly suggestive of a clean sample ($\Gamma\ll\Delta_0$) with a finite $c$, although it may be also consistent with a sample of extreme purity in the unitary limit. For all clean samples with a finite $c$ and the heat current along a point node a key feature of the temperature dependence of $\kappa/T$ is the non-monotonicity, the emergence of a characteristic  peak at low temperature~\cite{hirschfeld_consequences_1988,mishra2024thermal}. This peak disappears if a small gap opens in the nodal region.

Figure\,2c shows the calculated $\kappa/T$ (normalized as $\kappa\Delta_0/N_0v_F^2T$, where $N_0$ is the normal-state density of states, and $v_F$ is the Fermi velocity) as a function of $T/T_c$ for true point nodes ($\Delta_{\rm min} = 0$) and pseudo point nodes ($\Delta_{\rm min} = 0.1\Delta_0$), with {\boldmath $j$}$_Q$ applied parallel to the nodal direction. Here, we use representative values $\Gamma/\Delta_0=0.002$ and $c=0.15$. Technical details of the calculations and additional data demonstrating that this is a generic behavior are given in the Supplementary Information. Our samples exhibit negligibly small residual thermal conductivity. In the presence of point nodes along the $b$-axis, a low-temperature peak in $\kappa_b/T$ would be expected. Given that the quasiparticle contribution dominates the thermal conductivity, such a peak would be experimentally detectable. However, no indication of a peak is observed in either $\kappa_b/T$ or $\kappa_a/T$ \cite{suetsugu2024fully,hayes2025robust}, strongly suggesting a nodeless superconducting gap structure in the $ab$-plane. 

\begin{figure*}[t]
    \label{fig:TRSB}
    \includegraphics[clip,width=1.0\linewidth]{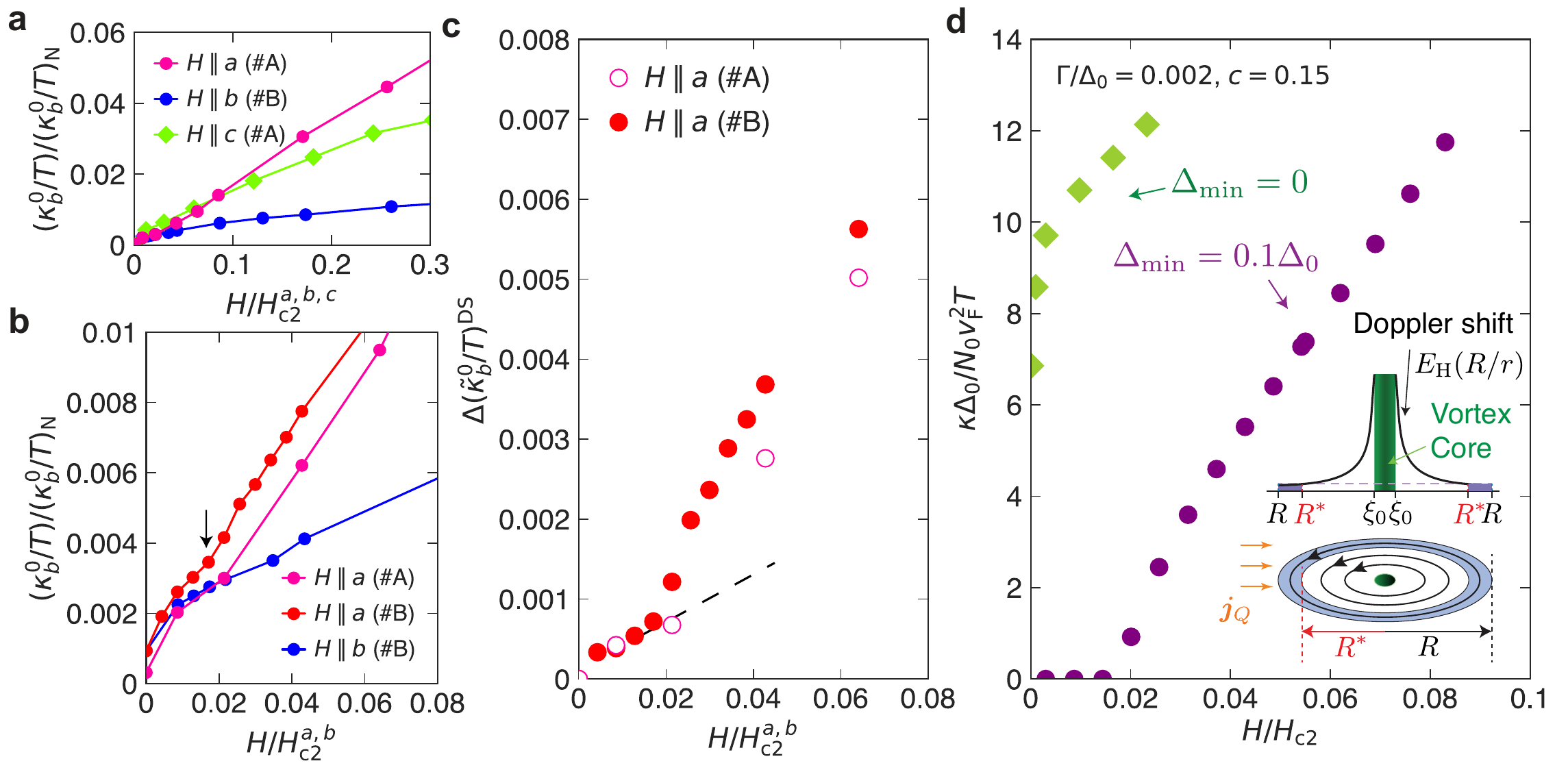}
    \caption{\textbf{
    Field-enhanced thermal conductivity in the $T \rightarrow 0$ limit: } \textbf{a--c}, The zero-temperature limit of the $b$-axis thermal conductivity divided by temperature $\kappa_0^b/T$, normalized by the normal state value $(\kappa_0^b/T)_N$, plotted as a function of magnetic field applied parallel to the $a$, $b$, and $c$ axes. The magnetic fields are normalized by the corresponding upper critical field $H_{c2}^{a,b,c}$.
\textbf{a}, $(\kappa_b^0/T)/(\kappa_b^0/T)_N$ as a function of $H/H_{c2}^{a,b,c}$ for {\boldmath $H$}$\parallel a$, $b$ and $c$. \textbf{b}, 
$(\kappa_b^0/T)/(\kappa_b^0/T)_N$ as a function of $H/H_{c2}^{a,b}$ for {\boldmath $H$}$\parallel a$ and $b$. For {\boldmath $H$}$\parallel a$, a kink is observed at $H^*/H_{c2}^a \approx 0.015$ (arrow), below which $(\kappa_b^0/T)/(\kappa_b^0/T)_N$ exhibits nearly isotropic behavior, whereas pronounced anisotropy develops above this characteristic threshold field. \textbf{c}, The red circles show the thermal conductivity difference $\Delta (\tilde{\kappa}_b^0/T)^{\mathrm{DS}} = [\Delta (\kappa_b^0/T)/(\kappa_b^0/T)_N]_{{\boldsymbol H} \parallel a} - [\Delta (\kappa_b^0/T)/(\kappa_b^0/T)_N]_{{\boldsymbol H} \parallel b}$ for \#B, which represents the difference between field configurations with and without the Doppler shift contribution.  For comparison, pink open circles represent the corresponding difference calculated by subtracting the \#B (${\boldsymbol H} \parallel b$) data from the \#A (${\boldsymbol H} \parallel a$) data. Both datasets exhibit good agreement. The dashed line is a guide to the eye. \textbf{d}, Magnetic field dependence of the residual thermal conductivity term for the true point nodes ($\Delta_{\rm min}=0$) and pseudo-nodes ($\Delta_{\rm min}=0.1\Delta_0$) in a model of a Fermi surface with a square cross-section in the $a$-$b$ plane and nodes/minima along the $b$ axis, as described in the Supplementary Information.  Inset: Schematic of Doppler shift effects in a model circular unit cell of the vortex lattice (radius $R$). Below the threshold field $E_{H}^\star=\Delta_{\rm min}$ the Doppler shift at distances $R>r>R^\star\sim R E_H/\Delta_{\rm min}$ is insufficient to generate unpaired quasiparticles (blue ring). For the heat current normal to the field this yields an effective thermally insulating barrier, preventing net heat conduction, see text for details.
}%
\end{figure*}

Thermal conductivity measurements with applied magnetic fields along three crystallographic orientations provide a more stringent and direct test for the absence of point nodes and the presence of a small gap. Magnetic fields induce circulating supercurrents around vortices characterized by the superfluid momentum $\boldsymbol{p}_s(\boldsymbol{r})$ in the plane normal to the field. 
The resulting Doppler shift in quasiparticle energy, $E_k \rightarrow E_k - \boldsymbol{v}_{\rm F} \cdot \boldsymbol{p}_s(\boldsymbol{r})$ (Figs.\,1d and e) selectively activates delocalized quasiparticles with the Fermi velocity aligned with $\boldsymbol{p}_s$. Quasiparticle thermal transport in the low-field regime of nodal superconductors is dominated by these quasiparticles, whose density of states exhibits characteristic field dependence determined by nodal topology: $N_{\text{del}}(\varepsilon) \propto \sqrt{H}$ for line nodes and $N_{\text{del}}(\varepsilon) \propto H|\ln H|$ for point nodes \cite{graf1996electronic,vekhter1999quasiparticle,adachi2007quasi}. Consequently, the thermal conductivity typically also varies as a power law of the field. Conversely, in full-gap type-II superconductors, where quasiparticles remain localized within vortex cores, while the density of states scales linearly with the number of vortices and hence the field, the thermal conductivity remains suppressed.

Critically, thermal conductivity in field sensitively probes nodal quasiparticles  when $\boldsymbol{v}_{\rm F}$ has components parallel to both $\boldsymbol{p}_s(\boldsymbol{r})$ and $\boldsymbol{j}_Q$, but is not affected by quasiparticles  with $\boldsymbol{v}_{\rm F} \perp \boldsymbol{p}_s(\boldsymbol{r})$ or $\boldsymbol{v}_{\rm F} \perp \boldsymbol{j}_Q$ (Fig.\,1e). In the presence of point nodes along the $b$ direction, applying $\boldsymbol{H}$ perpendicular to the $b$ axis ($\boldsymbol{H} \parallel a$ or $\boldsymbol{H} \parallel c$) would activate Doppler-shifted quasiparticles near the nodes due to a finite $\boldsymbol{p}_s \cdot \boldsymbol{v}_{\rm F}$ component for quasiparticles with $\boldsymbol{v}_{\rm F} \parallel \boldsymbol{j}_Q \parallel b$, leading to enhanced thermal conductivity. Conversely, for $\boldsymbol{H} \parallel b$, the supercurrent circulates in the $ac$-plane, resulting in $\boldsymbol{p}_s \cdot \boldsymbol{v}_{\rm F} = 0$ for heat-carrying quasiparticles along the $b$ direction; thus, $\kappa_b$ would be almost insensitive to such nodes (Fig.\,1d). This directional sensitivity enables spatial mapping of gap nodes on the Fermi surface. Importantly, even in full-gap superconductors with significant gap anisotropy, quasiparticles are excited at finite fields when the Doppler shift exceeds the minimum gap. Therefore, low-temperature and low-field measurements taken together yield definitive information about the gap structure.

Figures\,3a--c show the temperature dependence of $\kappa_b/T$ normalized by the normal-state value $(\kappa_b/T)_{\rm N}$ for three field orientations down to $\sim$50\,mK and for the field values up to 6\,T. At 0.3\,K, zero-field $\kappa_b/T$ exhibits strong suppression under a magnetic field of 1\,T, attributed to enhanced quasiparticle scattering by magnetic flux. Given that $\kappa_b^{\rm ph}$ is insensitive to magnetic fields, this pronounced suppression confirms the dominant quasiparticle contribution to the total thermal conductivity in the low temperature regime. This conclusion is further supported by the anisotropic field response of $\kappa_b/T$ for the fields along the $a$ and $c$ axes, since $\kappa_b^{\rm ph}$ remains isotropic with respect to field direction normal to the heat current.

 Figure\,4a shows $\kappa_b^0/T$ normalized by the normal-state value $(\kappa_b^0/T)_N$ under applied magnetic fields along three crystallographic directions, where $\kappa_b^0/T \equiv \lim_{T\to0}\kappa_b(T)/T$ is obtained by polynomial extrapolation to $T \to 0$. The magnetic fields are normalized by their respective upper critical fields to account for the strong $H_{c2}$ anisotropy: $\mu_0H_{c2}^a=11.7$\,T, $\mu_0H_{c2}^b=23$\,T, and $\mu_0H_{c2}^c=16.5$\,T \cite{wu2024enhanced}. The thermal conductivity exhibits pronounced anisotropic field dependence at higher fields,  with $\kappa_b/T$ showing the weakest growth with field $\boldsymbol{H} \parallel b$ despite the expected normal-state contributions from vortex cores in the $\boldsymbol{j}_Q \parallel \boldsymbol{H}$ configuration. This indicates that vortex-core conduction is essentially negligible. One possible explanation for this is that the discreteness of  Caroli--de Gennes--Matricon vortex core states reduces the contribution of the cores to the heat transport. Using $\Delta_0 \sim 2 T_{\rm c}\sim4$\,K and $\epsilon_{\rm F}\sim120$\,K \cite{eaton2024quasi} as the characteristic superconducting gap energy scale and the Fermi energy respectively, we estimate the core level spacing to be $\Delta_0^2/\epsilon_{\rm F} \sim 0.13$\,K, in agreement with Ref.~\cite{sharma_observation_2025}. The phase diagram of UTe$_2$ for $\boldsymbol{H} \parallel b$ is complex at high fields, with re-entrant superconductivity and other phases. However,  in the range where the measurements are done the field in the vortex core is still far below $H_{c2}$, and therefore emergence of additional order parameters, while possible, does not seem likely.
 
 The low-field results provide strong evidence for pseudo-point nodes, characterized by gap minima that remain finite rather than reaching zero.  The in-plane field response of $\kappa_b^0/T$ shown in Fig.\,4b offers critical insights into the detailed structure of the superconducting gap. The most notable feature is the apparent kink at a threshold field $H_{a}^*/H_{c2}^{a} \sim 0.015$ (arrow) in the field dependence of $\kappa_b^0/T$ for $\boldsymbol{H} \parallel a$ for both samples \#A and \#B. This kink at the threshold is more clearly observed in Fig.\,4c showing the difference between $\boldsymbol{H} \parallel \boldsymbol{j}_Q$ ($\boldsymbol{H} \parallel b$) and $\boldsymbol{H} \perp \boldsymbol{j}_Q$ ($\boldsymbol{H} \parallel a$) configurations. These results demonstrate that quasiparticle excitations undergo a qualitative change at this threshold field. Below $H_a^*$, $\kappa_b^0/T$ exhibits only weak anisotropy between field orientations. Above $H_a^*$, the in-plane response becomes strongly anisotropic, with $\boldsymbol{H} \perp \boldsymbol{j}_Q$  yielding substantially larger values than $\boldsymbol{H} \parallel \boldsymbol{j}_Q$. The perpendicular field geometry more efficiently generates Doppler-shifted quasiparticles with velocities aligned along the thermal gradient. Consequently, the absence of significant anisotropy below $H_a^*$ indicates that such Doppler-shifted quasiparticles contribute very little to thermal transport across the vortex lattice. This is consistent with a finite minimal gap.

Consider the geometry for fields normal to the heat current, as illustrated in Fig.\,4d. In the approximation where the vortex unit cell is a circle of radius $R\sim\xi_0\sqrt{H_{c2}/H}$ ($\xi_0$ is the coherence length),  the characteristic Doppler shift energy scale $E_H=a\Delta_0\sqrt{H/H_{c2}}$, with $a\sim 1$ is the {\it minimal } Doppler shift at the edge of the unit cell. The Doppler shift itself varies as $E_H(R/r)$  with the distance $r$ to the vortex core inside the unit cell. In the calculations below we took $a=0.72$, which is the value accepted in Ref.~\cite{hayes2025robust}, but the qualitative picture does not depend on the exact value of $a$.  To generate the unpaired quasiparticles in the absence of impurities, the Doppler shift needs to exceed the gap amplitude. Therefore in the presence of the minimal gap, such quasiparticles are generated everywhere in real space above the threshold field, $E_H^*=\Delta_{\rm min}$, or $H^*/H_{c2}=a^{-2}(\Delta_{\rm min}/\Delta_0)^2$. Below this field the unpaired quasiparticles only appear inside the circle of a radius $R^*=R (E_H/\Delta_{\rm min})<R$, while at $R^*<r<R$ the system is fully gapped (see Fig.\,4d).

Heat current normal to the vortices samples regions with all the Doppler shifts, and therefore the thermally insulating fully gapped region prevents heat conduction below $H^*$, yielding a sharp onset once this field is exceeded. This is shown in the main panel of Fig.\,4d, where we calculated the thermal conductivity following Ref.~\cite{kubert_quasiparticle_1998}, which took the  ``series'' resistor connection to model transport normal to the vortices. The threshold field is clearly seen at $H/H_{c2}\sim 0.015$ for the pseudo-nodal model assuming $\Delta_{\rm min}=0.1\Delta_0$ and is absent for true nodes. Very rapid rise of the thermal conductivity at low fields for $\Delta_{\rm min}=0$ is due to the square shape of the Fermi surface assumed in the model, see Supplementary Information for details. Therefore in this model the experimental results are consistent with  $\Delta_{\rm min}=0.1\Delta_0$ (Supplementary Fig.\,5), although the exact magnitude of the minimal gap depends on the choice of parameter $a$. In our model, as in Ref.~\cite{kubert_quasiparticle_1998}, the quasiparticles sample all available Doppler shift values along their path. We expect that beyond the semiclassical approach there will be finite, albeit small, heat transport across the classically forbidden regions, but expect that the kink at the threshold field will remain. 

Notably, the experimentally measured $\kappa/T$ exhibits a slight increase with applied field even below the threshold, in contrast to theoretical predictions. The discrepancy reflects limitations of the Doppler shift picture for transport. In the Doppler shift method one computes the local value of the transport coefficient. However,  if the range over which the corresponding correlation function decays encompasses regions of different superfluid velocities, the results are only qualitatively correct even at low fields. In particular, if the mean free path is comparable to or greater than the width of the ``forbidden'' region in the schematic of Fig.\,4d, the quasiparticles can traverse that region, and the heat current will be finite, albeit reduced. 

Quantum oscillation studies, which serve as a bulk probe of electronic structure, have demonstrated that UTe$_2$ possesses distinct cylindrical electron and hole Fermi surface sheets that are topologically separated, with no connectivity at Brillouin zone boundaries \cite{aoki2023haas}. Initially, quantum oscillation measurements suggested the presence of a 3D Fermi surface pocket \cite{broyles2023revealing}. However, subsequent high-resolution studies found no evidence for 3D Fermi surface sections \cite{eaton2024quasi} and demonstrated that the oscillations originate from quantum interference effects between quasi-2D sheets \cite{weinberger2024quantum,husstedt2025slow}. Based on these findings, we adopt a quasi-2D Fermi surface topology. In this topology, the gap function is nodeless due to the absence of Fermi surface sections extending along the $c$-axis direction, and the thermal conductivity under elevated fields reveals intrinsic gap anisotropy in 2D Fermi sheets.

Based on the observed fully gapped superconducting behavior in the \textit{ab}-plane, the order parameter is expected to belong to either the \textit{A}$_u$ or \textit{B}$_{1u}$ irreducible representation. In the \textit{B}$_{1u}$ state, the gap function exhibits point nodes along the \textit{c} axis; however, these remain experimentally inaccessible owing to the absence of Fermi-surface sections along this direction. Within the quasi-2D electronic structure, theoretical calculations suggest that UTe$_2$ likely realizes a topological crystalline superconducting state protected by time-reversal and crystal symmetries \cite{tei2023possible}. Both pairing states support Majorana edge modes at the same open boundaries perpendicular to the \textit{a} or \textit{b} axes. The distinction lies in their dispersion: a flat band for the \textit{B}$_{1u}$ state and a linearly dispersing mode for the \textit{A}$_u$ state. These theoretical predictions could be tested by orientation-resolved STM, although surface CDW effects may obscure intrinsic topological superconducting signatures.

 We next compare our conclusions to the reported results from bulk probes of the gap function, such as  nuclear magnetic resonance (NMR) and specific heat measurements. The observed suppression of NMR Knight shifts along all crystallographic axes appears consistent with \textit{A}$_u$ symmetry \cite{matsumura2023large}; however, this holds only when the $\boldsymbol{d}$-vector contains linear momentum terms proportional to $k_i$ (where $i = a, b, c$). However, $\boldsymbol{d}$-vectors incorporating higher-order momentum components, such as cubic terms $k_a k_b k_c$, produce finite contributions along all crystallographic directions, resulting in Knight shift suppression along all axes irrespective of the underlying gap symmetry. Consequently, the observed Knight shift suppression cannot unambiguously distinguish between \textit{A}$_u$, \textit{B}$_{1u}$, \textit{B}$_{2u}$, and \textit{B}$_{3u}$ representations, nor can it conclusively identify the fully gapped superconducting order.

Specific-heat measurements on UTe$_2$ exhibit power-law behavior, $C_{\text{el}} \propto T^n$ with $n \approx 3$ in zero field, attributed to quasiparticle excitations from point nodes~\cite{metz2019point,lee2025anisotropic}. The anisotropic field dependence of $C_{\text{el}}$ has been interpreted as evidence for point nodes oriented along the \textit{b}-axis~\cite{lee2025anisotropic}. However, the proposed \textit{b}-axis pseudo point node scenario consistently explains both behaviors (Supplementary Fig.\,6). The gap minima revealed by thermal conductivity measurements remain inaccessible to specific-heat measurements due to Schottky anomalies below $T \sim 0.3\,\mathrm{K}$, which are strongly enhanced under applied magnetic fields. As shown in the Supplementary Information, the $T$-dependence of the specific heat exhibits $C/T\propto T^2$ dependence above $0.3\,\mathrm{K}$ for pseudo point nodes with $\Delta_{\rm min}/\Delta_0\approx 0.1$. Magnetic field dependence of the specific heat is also unlikely to identify a small residual gap. While for the point nodes the expected field behavior is $C/T\propto -(H/H_{c2})\ln H/H_{c2}$, for a small gap it becomes $C/T\propto -(H/H_{c2})\ln \left[H/H_{c2}+(\Delta_{\rm min}/a\Delta_0)^2\right]$, and hence the substantial difference is only observed for $H\ll H^\star$, where such measurements are not possible. 

Finally, the emergence of a strongly anisotropic superconducting gap exhibiting pseudo point nodes in UTe$_2$ is highly unusual and represents a remarkable example of unconventional superconductivity. A possible mechanism may involve an accidental gap minimum arising from specific band structure characteristics. However, the formation of a pseudo point nodal gap characterized by a finite gap anisotropy ratio $\Delta_{\rm min}/\Delta_0 \sim 0.1$ would necessitate precise fine-tuning of both the pairing interaction strength and the degree of Fermi surface corrugation. Nevertheless, both the electron and hole Fermi surfaces are rather smooth, without any local corrugation that could produce point nodes\cite{eaton2024quasi}. Therefore, the Fermi surface geometry and topology of UTe$_2$ are relatively straightforward, and the accidental emergence of such an unconventional gap structure, while perhaps not theoretically precluded entirely, appears highly improbable. This suggests that the pseudo point node represents an anomalous behavior from the perspective of conventional pairing mechanisms.

The presence of pseudo-point nodes in UTe$_2$ imposes important constraints on the superconducting gap function and its formation mechanism. Non-unitary mixing states with broken time-reversal symmetry cannot readily explain this phenomenon. Indeed, non-unitary mixing of $B_{2u}$ symmetry, which has point nodes along the $b$-axis, with a small admixture of other symmetries (e.g., $B_{2u}$+i$\alpha A_u$ with $\alpha\ll 1$) merely shifts the node positions while preserving the nodes. To completely lift the point nodes, it is necessary to consider unitary mixing states where two order parameter components, such as $B_{2u}$+$\alpha A_u$, possess real relative phases. Alternatively, pseudo-point nodes can be interpreted as nodal structures that are quasi-protected by crystal symmetry but partially lifted by weak symmetry-breaking perturbations, possibly arising from spin-orbit coupling or weak interorbital hybridization. This results in the formation of a highly anisotropic superconducting gap with pronounced minima along specific crystallographic directions in momentum space. Further theoretical analysis and experimental verification are required to identify the dominant mechanism underlying pseudo-point node formation in UTe$_2$.

\section*{Methods}
High-quality single-crystal samples of UTe$_2$ were grown by the molten-salt method. Both crystals were cut to dimensions suitable for thermal-conductivity measurements, typically with length $\sim 360$\,{\textmu}m ($\parallel b$), width $\sim 370$\,{\textmu}m ($\parallel a$), and thickness $\sim 110$\,{\textmu}m ($\parallel c$). Thermal conductivity was measured using the standard steady-state method with one heater and two thermometers, applying a temperature gradient along the $b$ axis. The typical thermal gradient was less than 3\% of the bath temperature. To ensure good electrical and thermal contacts, four platinum wires were attached by spot welding. Electrical resistivity was measured \textit{in situ} within the same set-up used for thermal-conductivity measurements.

\section*{Acknowledgments}
We thank D. Aoki, A. Balatsky, S. Fujimoto, H. Kontani, S. Lin, T. Shibauchi, S. Suetsugu,  J. Thompson, and Y. Yanase for insightful discussions.
 Work at the Los Alamos National Laboratory was supported by the U.S. Department of Energy, Office of Basic Energy Sciences, Division of Materials Science and Engineering project ``Quantum Fluctuations in Narrow-band Systems''. S.H. acknowledges the Director's Postdoctoral Fellowship through the Laboratory Directed Research and Development Program at Los Alamos National Laboratory.

 \section*{Author contributions}
S.H. and Y.M. conceived the study. M.M.B., E.D.B and P.F.S.R. synthesized the samples. S.H., K.I., and R.M. performed thermal conductivity and resistivity measurements. S.H. analyzed the data, and I.V. performed theoretical calculations. All authors discussed the results. S.H., Y.M. and I.V.  prepared the manuscript with input from all authors.

\bibliography{refUTe2}

@article{broyles2023revealing,
  title={Revealing a 3D Fermi surface pocket and electron-hole tunneling in UTe$_2$ with quantum oscillations},
  author={Broyles, Christopher and Rehfuss, Zack and Siddiquee, Hasan and Zhu, Jiahui Althena and Zheng, Kaiwen and Nikolo, Martin and Graf, David and Singleton, John and Ran, Sheng},
  journal={Phys. Rev. Lett.},
  volume={131},
  number={3},
  pages={036501},
  year={2023},
  publisher={APS}
}

@article{eaton2024quasi,
  title={Quasi-2D Fermi surface in the anomalous superconductor UTe$_2$},
  author={Eaton, AG and Weinberger, TI and Popiel, NJM and Wu, Z and Hickey, AJ and Cabala, A and Posp{\'\i}{\v{s}}il, J and Prokle{\v{s}}ka, J and Haidamak, T and Bastien, G and others},
  journal={Nat. Commun.},
  volume={15},
  number={1},
  pages={223},
  year={2024},
  publisher={Nature Publishing Group UK London}
}

@article{weinberger2024quantum,
  title={Quantum interference between quasi-2D Fermi surface sheets in UTe$_2$},
  author={Weinberger, TI and Wu, Z and Graf, DE and Skourski, Y and Cabala, A and Posp{\'\i}{\v{s}}il, J and Prokle{\v{s}}ka, J and Haidamak, T and Bastien, G and Sechovsk{\`y}, V and others},
  journal={Phys. Rev. Lett.},
  volume={132},
  number={26},
  pages={266503},
  year={2024},
  publisher={APS}
}

@article{ran2019nearly,
  title={Nearly ferromagnetic spin-triplet superconductivity},
  author={Ran, Sheng and Eckberg, Chris and Ding, Qing-Ping and Furukawa, Yuji and Metz, Tristin and Saha, Shanta R and Liu, I-Lin and Zic, Mark and Kim, Hyunsoo and Paglione, Johnpierre and others},
  journal={Science},
  volume={365},
  number={6454},
  pages={684--687},
  year={2019},
  publisher={American Association for the Advancement of Science}
}

@article{saxena2000superconductivity,
  title={{Superconductivity on the border of itinerant-electron ferromagnetism in UGe$ _2$}},
  author={Saxena, SS and Agarwal, P and Ahilan, K and Grosche, FM and Haselwimmer, RKW and Steiner, MJ and Pugh, E and Walker, IR and Julian, SR and Monthoux, P and others},
  journal={Nature},
  volume={406},
  number={6796},
  pages={587--592},
  year={2000},
  publisher={Nature Publishing Group}
}

@article{aoki2001coexistence,
  title={{Coexistence of superconductivity and ferromagnetism in URhGe}},
  author={Aoki, Dai and Huxley, Andrew and Ressouche, Eric and Braithwaite, Daniel and Flouquet, Jacques and Brison, Jean-Pascal and Lhotel, Elsa and Paulsen, Carley},
  journal={Nature},
  volume={413},
  number={6856},
  pages={613--616},
  year={2001},
  publisher={Nature Publishing Group}
}

@article{PhysRevLett.99.067006,
  title = {{Superconductivity on the Border of Weak Itinerant Ferromagnetism in UCoGe}},
  author = {Huy, N. T. and Gasparini, A. and de Nijs, D. E. and Huang, Y. and Klaasse, J. C. P. and Gortenmulder, T. and de Visser, A. and Hamann, A. and G\"orlach, T. and L\"ohneysen, H. v.},
  journal = {Phys. Rev. Lett.},
  volume = {99},
  issue = {6},
  pages = {067006},
  numpages = {4},
  year = {2007},
  month = {Aug},
  publisher = {American Physical Society},
  doi = {10.1103/PhysRevLett.99.067006}
}

@article{ran2019extreme,
  title={Extreme magnetic field-boosted superconductivity},
  author={Ran, Sheng and Liu, I-Lin and Eo, Yun Suk and Campbell, Daniel J and Neves, Paul M and Fuhrman, Wesley T and Saha, Shanta R and Eckberg, Christopher and Kim, Hyunsoo and Graf, David and others},
  journal={Nat. Phys.},
  volume={15},
  number={12},
  pages={1250--1254},
  year={2019},
  publisher={Nature Publishing Group}
}

@article{wu2025quantum,
  title={A quantum critical line bounds the high field metamagnetic transition surface in {UTe$_2$}},
  author={Wu, Zheyu and Weinberger, TI and Hickey, AJ and Chichinadze, DV and Shaffer, D and Cabala, A and Chen, H and Long, M and Brumm, TJ and Xie, W and others},
  journal={Phys. Rev. X},
  volume={15},
  number={2},
  pages={021019},
  year={2025},
  publisher={APS}
}

@article{knafo2021comparison,
  title={{Comparison of two superconducting phases induced by a magnetic field in UTe$ _2$}},
  author={Knafo, William and Nardone, M and Vali{\v{s}}ka, M and Zitouni, A and Lapertot, G and Aoki, D and Knebel, G and Braithwaite, D},
  journal={Commun. Phys.},
  volume={4},
  number={1},
  pages={40},
  year={2021},
  publisher={Nature Publishing Group}
}

@article{girod2022thermodynamic,
  title = {{Thermodynamic and electrical transport properties of ${\mathrm{UTe}}_{2}$ under uniaxial stress}},
  author = {Girod, Cl{\'e}ment and Stevens, Callum R. and Huxley, Andrew and Bauer, Eric D. and Santos, Frederico B. and Thompson, Joe D. and Fernandes, Rafael M. and Zhu, Jian-Xin and Ronning, Filip and Rosa, Priscila F. S. and Thomas, Sean M.},
  journal = {Phys. Rev. B},
  volume = {106},
  issue = {12},
  pages = {L121101},
  numpages = {5},
  year = {2022},
  month = {Sep},
  publisher = {American Physical Society},
  doi = {10.1103/PhysRevB.106.L121101}
}

@article{nayak2008non,
  title={Non-Abelian anyons and topological quantum computation},
  author={Nayak, Chetan and Simon, Steven H and Stern, Ady and Freedman, Michael and Das Sarma, Sankar},
  journal={Rev. Mod. Phys.},
  volume={80},
  number={3},
  pages={1083--1159},
  year={2008},
  publisher={APS}
}

@article{bae2021anomalous,
  title={{Anomalous normal fluid response in a chiral superconductor UTe$ _2$}},
  author={Bae, Seokjin and Kim, Hyunsoo and Eo, Yun Suk and Ran, Sheng and Liu, I-lin and Fuhrman, Wesley T and Paglione, Johnpierre and Butch, Nicholas P and Anlage, Steven M and others},
  journal={Nat. Commun.},
  volume={12},
  number={1},
  pages={1--5},
  year={2021},
  publisher={Nature Publishing Group}
}

@article{ishihara2021chiral,
  title={{Chiral superconductivity in UTe$ _2$ probed by anisotropic low-energy excitations}},
  author={Ishihara, Kota and Roppongi, Masaki and Kobayashi, Masayuki and Imamura, Kumpei and Mizukami, Yuta and Sakai, Hironori and Opletal, Petr and Tokiwa, Yoshifumi and Haga, Yoshinori and Hashimoto, Kenichiro and Shibauchi, Takasada},
  journal={Nat. Commun.},
  volume={14},
  number={1},
  pages={2966},
  year={2023},
  publisher={Nature Publishing Group UK London}
}

@article{PhysRevLett.123.217001,
  title = {{Insulator-Metal Transition and Topological Superconductivity in ${\mathrm{UTe}}_{2}$ from a First-Principles Calculation}},
  author = {Ishizuka, Jun and Sumita, Shuntaro and Daido, Akito and Yanase, Youichi},
  journal = {Phys. Rev. Lett.},
  volume = {123},
  issue = {21},
  pages = {217001},
  numpages = {6},
  year = {2019},
  month = {Nov},
  publisher = {American Physical Society},
  doi = {10.1103/PhysRevLett.123.217001}
}

@article{tei2023possible,
  title={Possible realization of topological crystalline superconductivity with time-reversal symmetry in UTe$_2$},
  author={Tei, Jushin and Mizushima, Takeshi and Fujimoto, Satoshi},
  journal={Phys. Rev. B},
  volume={107},
  number={14},
  pages={144517},
  year={2023},
  publisher={APS}
}

@article{aoki2023haas,
  title={de Haas--van Alphen Oscillations for the Field Along $c$-axis in UTe$_2$},
  author={Aoki, Dai and Sheikin, Ilya and McCollam, Alix and Ishizuka, Jun and Yanase, Youichi and Lapertot, Gerard and Flouquet, Jacques and Knebel, Georg},
  journal={J. Phys. Soc. Jpn.},
  volume={92},
  number={6},
  pages={065002},
  year={2023},
  publisher={The Physical Society of Japan}
}

@article{husstedt2025slow,
  title={Slow magnetic quantum oscillations in the c-axis magnetoresistance of UTe$_2$},
  author={Husstedt, Freya and Kimata, Motoi and Thadathil, Sajal Naduvile and Schwarze, Beat Valentin and K{\"o}nig, Markus and Lapertot, Gerard and Brison, Jean-Pascal and Knebel, Georg and Aoki, Dai and Wosnitza, J and others},
  journal={Phys. Rev. B},
  volume={111},
  number={23},
  pages={235131},
  year={2025},
  publisher={APS}
}

@article{PhysRevMaterials.6.073401,
  title = {{Single crystal growth of superconducting ${\mathrm{UTe}}_{2}$ by molten salt flux method}},
  author = {Sakai, H. and Opletal, P. and Tokiwa, Y. and Yamamoto, E. and Tokunaga, Y. and Kambe, S. and Haga, Y.},
  journal = {Phys. Rev. Mater.},
  volume = {6},
  issue = {7},
  pages = {073401},
  numpages = {10},
  year = {2022},
  month = {Jul},
  publisher = {American Physical Society},
  doi = {10.1103/PhysRevMaterials.6.073401}
}

@article{ajeesh2023fate,
  title={Fate of time-reversal symmetry breaking in UTe$_2$},
  author={Ajeesh, MO and Bordelon, M and Girod, C and Mishra, Sanu and Ronning, Filip and Bauer, Eric Dietzgen and Maiorov, B and Thompson, Joe David and Rosa, PFS and Thomas, Sean Michael},
  journal={Phys. Rev. X},
  volume={13},
  number={4},
  pages={041019},
  year={2023},
  publisher={APS}
}

@article{aishwarya2023magnetic,
  title={Magnetic-field-sensitive charge density waves in the superconductor UTe$_2$},
  author={Aishwarya, Anuva and May-Mann, Julian and Raghavan, Arjun and Nie, Laimei and Romanelli, Marisa and Ran, Sheng and Saha, Shanta R and Paglione, Johnpierre and Butch, Nicholas P and Fradkin, Eduardo and others},
  journal={Nature},
  volume={618},
  number={7967},
  pages={928--933},
  year={2023},
  publisher={Nature Publishing Group UK London}
}

@article{talavera2025surface,
  title={Surface charge density wave in UTe$_2$},
  author={Talavera, Pablo Garc{\'\i}a and Velasco, Miguel {\'A}gueda and Shimizu, Makoto and Wu, Beilun and Knebel, Georg and Patino, Midori Amano and Lapertot, Gerard and Flouquet, Jacques and Brison, Jean Pascal and Aoki, Dai and others},
  journal={arXiv:2504.12505},
  year={2025}
}

@article{iguchi2022microscopic,
  title = {{Microscopic Imaging Homogeneous and Single Phase Superfluid Density in ${\mathrm{UTe}}_{2}$}},
  author = {Iguchi, Yusuke and Man, Huiyuan and Thomas, S. M. and Ronning, Filip and Rosa, Priscila F. S. and Moler, Kathryn A.},
  journal = {Phys. Rev. Lett.},
  volume = {130},
  issue = {19},
  pages = {196003},
  numpages = {6},
  year = {2023},
  month = {May},
  publisher = {American Physical Society},
  doi = {10.1103/PhysRevLett.130.196003},
}

@article{lee2025anisotropic,
  title={Anisotropic field-induced changes in the superconducting order parameter of UTe$_2$},
  author={Lee, Sangyun and Woods, Andrew J and Rosa, PFS and Thomas, SM and Bauer, ED and Lin, Shi-Zeng and Movshovich, R},
  journal={Phys. Rev. Research},
  volume={7},
  number={2},
  pages={L022053},
  year={2025},
  publisher={APS}
}

@article{theuss2024single,
  title={Single-component superconductivity in UTe$_2$ at ambient pressure},
  author={Theuss, Florian and Shragai, Avi and Grissonnanche, Gael and Hayes, Ian M and Saha, Shanta R and Eo, Yun Suk and Suarez, Alonso and Shishidou, Tatsuya and Butch, Nicholas P and Paglione, Johnpierre and others},
  journal={Nature Physics},
  volume={20},
  number={7},
  pages={1124--1130},
  year={2024},
  publisher={Nature Publishing Group UK London}
}

@article{matsuda2006nodal,
  title={Nodal structure of unconventional superconductors probed by angle resolved thermaltransport measurements},
  author={Matsuda, Y and Izawa, K and Vekhter, I},
  journal={J. Phys. Condens. Matter},
  volume={18},
  number={44},
  pages={R705},
  year={2006},
  publisher={IOP Publishing}
}

@article{mishra2024thermal,
  title={Thermal conductivity of nonunitary triplet superconductors: application to UTe$_2$},
  author={Mishra, Vivek and Wang, Ge and Hirschfeld, PJ},
  journal={Front. Phys.},
  volume={12},
  pages={1397524},
  year={2024},
  publisher={Frontiers Media SA}
}

@article{hayes2025robust,
  title={Robust nodal behavior in the thermal conductivity of superconducting UTe$_2$},
  author={Hayes, Ian M and Metz, Tristin E and Frank, Corey E and Saha, Shanta R and Butch, Nicholas P and Mishra, Vivek and Hirschfeld, Peter J and Paglione, Johnpierre},
  journal={Phys. Rev. X},
  volume={15},
  number={2},
  pages={021029},
  year={2025},
  publisher={APS}
}

@article{suetsugu2024fully,
  title={Fully gapped pairing state in spin-triplet superconductor UTe$_2$},
  author={Suetsugu, Shota and Shimomura, Masaki and Kamimura, Masashi and Asaba, Tomoya and Asaeda, Hiroto and Kosuge, Yuki and Sekino, Yuki and Ikemori, Shun and Kasahara, Yuichi and Kohsaka, Yuhki and others},
  journal={Sci. Adv.},
  volume={10},
  number={6},
  pages={eadk3772},
  year={2024},
  publisher={American Association for the Advancement of Science}
}

@article{matsumura2023large,
  title={{Large Reduction in the $a$-axis Knight Shift on UTe$_2$ with $T_c$= 2.1 K}},
  author={Matsumura, Hiroki and Fujibayashi, Hiroki and Kinjo, Katsuki and Kitagawa, Shunsaku and Ishida, Kenji and Tokunaga, Yo and Sakai, Hironori and Kambe, Shinsaku and Nakamura, Ai and Shimizu, Yusei and others},
  journal={J. Phys. Soc. Jpn.},
  volume={92},
  number={6},
  pages={063701},
  year={2023},
  publisher={The Physical Society of Japan}
}

@article{metz2019point,
  title={Point-node gap structure of the spin-triplet superconductor UTe$_2$},
  author={Metz, Tristin and Bae, Seokjin and Ran, Sheng and Liu, I-Lin and Eo, Yun Suk and Fuhrman, Wesley T and Agterberg, Daniel F and Anlage, Steven M and Butch, Nicholas P and Paglione, Johnpierre},
  journal={Phys. Rev. B},
  volume={100},
  number={22},
  pages={220504},
  year={2019},
  publisher={APS}
}

@article{graf1996electronic,
  title={Electronic thermal conductivity and the Wiedemann-Franz law for unconventional superconductors},
  author={Graf, Matthias J and Yip, SK and Sauls, JA and Rainer, D},
  journal={Phys. Rev. B},
  volume={53},
  number={22},
  pages={15147},
  year={1996},
  publisher={APS}
}

@article{kengle2024absence,
  title={Absence of bulk charge density wave order in the normal state of UTe$_2$},
  author={Kengle, Caitlin S and Vonka, Jakub and Francoual, Sonia and Chang, Johan and Abbamonte, Peter and Janoschek, Marc and Rosa, Priscila FS and Simeth, Wolfgang},
  journal={Nat. Commun.},
  volume={15},
  number={1},
  pages={9713},
  year={2024},
  publisher={Nature Publishing Group UK London}
}

@article{vekhter1999quasiparticle,
  title={Quasiparticle thermal conductivity in the vortex state of high-$T$$_{\rm c}$ cuprates},
  author={Vekhter, I and Houghton, A},
  journal={Phys. Rev. Lett.},
  volume={83},
  number={22},
  pages={4626},
  year={1999},
  publisher={APS}
}

@article{wu2024enhanced,
  title={Enhanced triplet superconductivity in next-generation ultraclean UTe$_2$},
  author={Wu, Z and Weinberger, TI and Chen, J and Cabala, A and Chichinadze, DV and Shaffer, D and Posp{\'\i}{\v{s}}il, J and Prokle{\v{s}}ka, J and Haidamak, T and Bastien, G and others},
  journal={Proc. Natl. Acad. Sci. USA},
  volume={121},
  number={37},
  pages={e2403067121},
  year={2024},
  publisher={National Academy of Sciences}
}

@article{adachi2007quasi,
  title={Quasi-classical calculation of the mixed-state thermal conductivity in $s$-and $d$-wave superconductors},
  author={Adachi, Hiroto and Miranovic, Predrag and Ichioka, Masanori and Machida, Kazushige},
  journal={J. Phys. Soc. Jpn.},
  volume={76},
  number={6},
  pages={064708--064708},
  year={2007},
  publisher={The Physical Society of Japan}
}

@article{gu2025pair,
  title={Pair wave function symmetry in UTe$_2$ from zero-energy surface state visualization},
  author={Gu, Qiangqiang and Wang, Shuqiu and Carroll, Joseph P and Zhussupbekov, Kuanysh and Broyles, Christopher and Ran, Sheng and Butch, Nicholas P and Horn, Jarryd A and Saha, Shanta and Paglione, Johnpierre and others},
  journal={Science},
  volume={388},
  number={6750},
  pages={938--944},
  year={2025},
  publisher={American Association for the Advancement of Science}
}

@article{li2025observation,
  title={Observation of odd-parity superconductivity in UTe$_2$},
  author={Li, Zixuan and Moir, Camilla M and McKee, Nathan J and Lee-Wong, Eric and Baumbach, Ryan E and Maple, M Brian and Liu, Ying},
  journal={Proc. Natl. Acad. Sci. USA},
  volume={122},
  number={13},
  pages={e2419734122},
  year={2025},
  publisher={National Academy of Sciences}
}

@article{yoon2024probing,
  title={Probing p-wave superconductivity in UTe$_2$ via point-contact junctions},
  author={Yoon, Hyeok and Eo, Yun Suk and Park, Jihun and Horn, Jarryd A and Dorman, Ryan G and Saha, Shanta R and Hayes, Ian M and Takeuchi, Ichiro and Brydon, Philip MR and Paglione, Johnpierre},
  journal={npj Quantum Mater.},
  volume={9},
  number={1},
  pages={91},
  year={2024},
  publisher={Nature Publishing Group UK London}
}

@article{braithwaite2019multiple,
  title={Multiple superconducting phases in a nearly ferromagnetic system},
  author={Braithwaite, Daniel and Vali{\v{s}}ka, M and Knebel, G and Lapertot, G and Brison, J-P and Pourret, A and Zhitomirsky, ME and Flouquet, J and Honda, F and Aoki, D},
  journal={Commun. Phys.},
  volume={2},
  number={1},
  pages={147},
  year={2019},
  publisher={Nature Publishing Group UK London}
}

@article{thomas2020evidence,
  title={Evidence for a pressure-induced antiferromagnetic quantum critical point in intermediate-valence UTe{$_2$}},
  author={Thomas, SM and Santos, FB and Christensen, MH and Asaba, T and Ronning, F and Thompson, JD and Bauer, ED and Fernandes, RM and Fabbris, G and Rosa, PFS},
  journal={Sci. Adv.},
  volume={6},
  number={42},
  pages={eabc8709},
  year={2020},
  publisher={American Association for the Advancement of Science}
}

@article{azari2023absence,
  title={Absence of spontaneous magnetic fields due to time-reversal symmetry breaking in bulk superconducting {UTe$_2$}},
  author={Azari, N and Yakovlev, M and Rye, N and Dunsiger, SR and Sundar, S and Bordelon, MM and Thomas, SM and Thompson, JD and Rosa, PFS and Sonier, JE},
  journal={Phys. Rev. Lett.},
  volume={131},
  number={22},
  pages={226504},
  year={2023},
  publisher={APS}
}

@article{hirschfeld_consequences_1988,
    title = {Consequences of resonant impurity scattering in anisotropic superconductors: {Thermal} and spin relaxation properties},
    volume = {37},
    shorttitle = {Consequences of resonant impurity scattering in anisotropic superconductors},
    doi = {10.1103/PhysRevB.37.83},
    number = {1},
    urldate = {2025-09-04},
    journal = {Phys. Rev. B},
    author = {Hirschfeld, P. J. and Wölfle, P. and Einzel, D.},
    month = jan,
    year = {1988},
    pages = {83--97},
}

@article{kubert_quasiparticle_1998,
    title = {Quasiparticle {Transport} {Properties} of {$d$}-{Wave} {Superconductors} in the {Vortex} {State}},
    volume = {80},
    doi = {10.1103/PhysRevLett.80.4963},
    number = {22},
    urldate = {2025-10-04},
    journal = {Phys. Rev. Lett.},
    author = {Kübert, C. and Hirschfeld, P. J.},
    month = jun,
    year = {1998},
    pages = {4963--4966},
}

@article{sharma_observation_2025,
    title = {Observation of {Persistent} {Zero} {Modes} and {Superconducting} {Vortex} {Doublets} in {UTe$_2$}},
    volume = {19},
    issn = {1936-0851},
    doi = {10.1021/acsnano.5c08406},
    number = {35},
    urldate = {2025-10-15},
    journal = {ACS Nano},
    author = {Sharma, Nileema and Toole, Matthew and McKenzie, James and Cheng, Fangjun and Bordelon, Mitchell M. and Thomas, Sean M. and Rosa, Priscila F. S. and Hsu, Yi-Ting and Liu, Xiaolong},
    month = sep,
    year = {2025},
    pages = {31539--31550},
}

@article{aoki2022unconventional,
  title={Unconventional superconductivity in UTe{$_2$}},
  author={Aoki, Dai and Brison, Jean-Pascal and Flouquet, Jacques and Ishida, Kenji and Knebel, Georg and Tokunaga, Y and Yanase, Youichi},
  journal={J. Phys. Condens. Matter},
  volume={34},
  number={24},
  pages={243002},
  year={2022},
  publisher={IOP Publishing}
}

@article{xu2019quasi,
  title={Quasi-two-dimensional Fermi surfaces and unitary spin-triplet pairing in the heavy fermion superconductor UTe{$_2$}},
  author={Xu, Yuanji and Sheng, Yutao and Yang, Yi-feng},
  journal={Phys. Rev. Lett.},
  volume={123},
  number={21},
  pages={217002},
  year={2019},
  publisher={APS}
}

\end{document}